\documentclass[12pt]{article}
\usepackage{amssymb,amsmath,amscd,amsthm}
\usepackage[cp1251]{inputenc}
\usepackage[T2A]{fontenc}
\usepackage{pstricks}
\usepackage{graphicx,psfrag}

\newtheorem{theorem}{Theorem}[section]

\newtheorem{assumption}[theorem]{Assumption}

\newtheorem{proposition}[theorem]{Proposition}

\date{}

\begin{document}

\title{Remarks on interior transmission eigenvalues, Weyl formula and branching billiards}

\author{ E.Lakshtanov\thanks{Department of Mathematics, Aveiro University, Aveiro 3810, Portugal.  This work was supported by {\it FEDER} funds through {\it COMPETE}--Operational Programme Factors of Competitiveness (``Programa Operacional Factores de Competitividade'') and by Portuguese funds through the {\it Center for Research and Development in Mathematics and Applications} (University of Aveiro) and the Portuguese Foundation for Science and Technology (``FCT--Fund\c{c}\~{a}o para a Ci\^{e}ncia e a Tecnologia''), within project PEst-C/MAT/UI4106/2011 with COMPETE number FCOMP-01-0124-FEDER-022690, and by the FCT research project
PTDC/MAT/113470/2009 (lakshtanov@rambler.ru).} \and
 B.Vainberg\thanks{Department
of Mathematics and Statistics, University of North Carolina,
Charlotte, NC 28223, USA. The work was partially supported  by the NSF grant DMS-1008132 (brvainbe@uncc.edu).}}

\maketitle

\begin{abstract}
The paper contains the Weyl formula for the counting function of the interior transmission problem when the latter is parameter-elliptic. Branching billiard trajectories are constructed, and the second term of the Weyl asymptotics is estimated from above under some conditions on the set of periodic billiard trajectories.
\end{abstract}


\section{Interior transmission eigenvalues.}
Let $\mathcal O\in R^d$ be an open bounded domain with $C^\infty$ boundary $\partial O$.
The classical Weyl formula
\begin{equation}\label{classic}
N(\lambda) \sim \frac{V(\mathcal O)\omega_d}{(2\pi)^d} \lambda^{d/2}, \quad  \lambda \rightarrow \infty,
\end{equation}
for the counting function $N(\lambda)$ (number of eigenvalues whose absolute values do not exceed $\lambda$) is well known for the Dirichlet or Neumann Laplacian in $\mathcal O$. Here $V(\mathcal O)$ is the volume of the domain and $\omega_d$ is the volume of the unit ball in $R^d$.  This paper concerns the Weyl formula for the interior transmission eigenvalues (ITE) which were introduced by A.Kirsch in \cite{Kir} in connection with an inverse scattering problem for the reduced wave equation and further studied by D.Colton
and P.Monk \cite{base}. The anisotropic interior transmission problem (discussed below) was introduced in \cite{col}, see the review \cite{cps} for more references.

Let us recall the definition of ITE.
The values of $\lambda  \in \mathbb C$ for which the following problem
\begin{equation}\label{Anone}
\begin{array}{l}
-\Delta u - \lambda u =0, \quad \quad  \quad  \quad x \in \mathcal O, \quad u\in H^2(\mathcal O),\\
-\nabla A \nabla v - \lambda  n(x)v =0, \quad x \in \mathcal O, \quad v\in H^2(\mathcal O),
\end{array}
\end{equation}
\begin{equation}\label{Antwo}
\begin{array}{l}
u-v=0, \quad \quad x \in \partial \mathcal O, \\
\frac{\partial u}{\partial \nu} - \frac{\partial v}{\partial \nu_A}=0, \quad x \in \partial \mathcal O.
\end{array}
\end{equation}
 has a non-trivial solution are called the \textit{interior transmission eigenvalues}.
 Here $H^{2}(\mathcal O)$ is the Sobolev space, $A(x),~x\in \overline{\mathcal O}$, is a smooth symmetric elliptic ($A=A^t>0$) matrix with real-valued entries, $n(x)$ is  a smooth function, $\nu$ is the outward normal to $\partial\mathcal O$, and the co-normal derivative is defined as follows
$$
\frac{\partial } {\partial \nu_A}v =\nu \cdot A \nabla v.
$$
 We will mostly be concerned with the cases $d=2,3$, but all the results below can be automatically carried over to any dimension $d$.

There are many papers on the Weyl formula for general elliptic boundary value problems (see review \cite{A1} for references). The particular feature of the problem under consideration is that it is not symmetric.  The formally conjugate problem has different boundary conditions:  \begin{equation}\label{AntwoC}
\begin{array}{l}
u+v=0, \quad x \in \partial \mathcal O, \\
\frac{\partial u}{\partial \nu} + \frac{\partial v}{\partial \nu_A}=0, \quad x \in \partial \mathcal O.
\end{array}
\end{equation}
The spectrum of the problem (\ref{Anone}),(\ref{Antwo}) is not always discrete. Examples when the eigenvalues fill the whole complex plane can be found in \cite{LV4}. It was also shown there that ITE form a discrete set when the problem is parameter-elliptic. Conditions for parameter-ellipticity were also described in \cite{LV4}, and they will be formulated below in the case of real $A$ and $n$. This paper contains a justification of the Weyl formula for ITE when the problem (\ref{Anone}),(\ref{Antwo}) is parameter-elliptic.  In particular, this implies that the set of ITE is infinite for general parameter-elliptic problems. The infiniteness of the set of ITE was known in many cases, see \cite{cak}, \cite{cak1} and references there.
When $A=I\!d$, some estimates on $N(\lambda)$ for ITE can be found in  \cite{ss},\cite{tsm}.
Before we proceed with the main result on the Weyl formula, we would like to show that ITE play the same role for the transmission  scattering problem as the eigenvalues of the Dirichlet or Neumann Laplacian play for the scattering by an obstacle with the corresponding (Dirichlet or Neumann) boundary conditions.


A connection between the counting function $N(\lambda)$ for the Dirichlet or Neumann Laplacian and the total scattering phase (which is defined as $-\arg \text{det} S(k)$, where $S(k)$ is the scattering matrix) was established in 1978, see \cite{mr}, \cite{jk}. It was shown that
$$
-\frac{1}{2\pi}\arg\text{det} S(k) = N(\lambda)(1+O(\lambda^{-\varepsilon})), \quad \varepsilon >0, ~ \lambda=k^2 \rightarrow \infty.
$$
This formula suggests that $N(\lambda)$ can perhaps be interpreted as the rotation number of the scattering phase (number of rotations of  $\exp(i ~\mbox{arg det}S(k))$ on the unit circle when $k\to\infty$).  At present, there is a deeper understanding of the relation between $N(\lambda)$ and the rotation number.

Recall that the operator $S(k),~k>0,$ is unitary and its eigenvalues $z=z_j(k)$ belong to the unit circle. Consider two conditions.

A.  $-k^2$ is an eigenvalue of the Dirichlet or Neumann Laplacian.

B. $z=1$ is an eigenvalue of S-matrix ($S=S(k)$) for the scattering problem with the same boundary condition and $k>0$.

It is easy to see that B implies A. Indeed, the unitary matrix $S(k)$ has the form $S(k)=I+\frac{ik}{2\pi}F$, where $F$ is the integral operator on $L^2(S^{d-1})$ ($S^{d-1}$ is the unit sphere) whose kernel is the scattering amplitude. If $S\mu(\theta)=\mu(\theta),~\theta\in S^{d-1}$, then $F\mu =0$. Let $\psi(k,\omega,x)=e^{ik(\omega,x)}+\psi_{\text{sc}},~ \omega\in S^{d-1}$, be the solution of the scattering problem. Consider function $u(k,x)=\int_{S^{d-1}}\psi\mu(\omega)dS_\mu$. Function $u$ satisfies the homogeneous boundary condition since $\psi$ satisfies it. Since $F\mu =0$, the outgoing wave $\int_{S^{d-1}}\psi\mu(\omega)dS_\mu$ has zero amplitude, and therefore it is equal to zero identically. Thus $u(k,x)=\int_{S^{d-1}}e^{ik(\omega,x)}\mu(\omega)dS_\mu$. Obviously, this function satisfies the Helmholtz equation in the whole space, i.e., $u$ is an eigenfunction of the interior problem with the eigenvalue $-k^2$.

This implication from B to A was first noted in papers on inverse scattering problem (see  \cite{KC1984}) were slightly different terminology was used: injectivity of the far field operator $F$ implies A. In terms of $S$-matrix, it can be found in  \cite{DorSm}.

The inverse implication (from A to B) holds in the case of a ball, but not for general domains, see \cite{EP}. However, for general domains $\mathcal O$, B implies A in a weaker sense: $-k_0^2$ is an eigenvalue of the Dirichlet Laplacian if and only if there exists an analytic in $k\in(k_0-\delta,k_0),~\delta>0,$ eigenvalue $z=z_i(k)$ of the $S$-matrix such that
\begin{equation}\label{eee2}
\lim_{k \rightarrow k_0-0} \mbox{arg}  z_i(k)=2\pi\!+\!0.
  \end{equation}

In connection with the last relation, let us note that the eigenvalues $z=\{z_j(k)\}$ of the $S$-matrix $S(k)$ form an analytic manifold while $z\neq 1$. Point $z=1$ is an essential point of the spectrum of the operator $S(k)$ (limiting point for the set of eigenvalues) and the structure of the manifold in a neighborhood of this point is much more complicated.  In particular, it may happen that an eigenvalue approaches $z=1$ as $k\to k_0$, but $z=1$ is not an eigenvalue of $S(k_0)$. It is also essential that we have  $k\to k_0\!-\!0$ in (\ref{eee2}), but not $k\to k_0\!+\!0$.

The study of the connection between properties A and B was initiated by E. Doron, U. Smilansky,  \cite{DorSm}. Relation (\ref{eee2}) was justified by J. Eckmann and C. Pillett in \cite{EP} in the case of the Dirichlet boundary condition. Paper \cite{EP2} contains a similar result for the Neumann boundary condition.

All the  relations between the Dirichlet/Neumann problem and scattering problem mentioned above can be easily reformulated for ITE and the (exterior) transmission scattering problem. The latter problem is stated as follows.
\begin{equation}\label{AnoneE}
\begin{array}{l}
-\Delta u -\lambda u =0, \quad x \in R^d\backslash\mathcal O, \quad u=e^{ik(x,\omega)}+\psi_{\text{sc}}(x,k,\omega),\\
-\nabla A \nabla v - \lambda  n(x)v =0, \quad x \in \mathcal O,
\end{array}
\end{equation}
\begin{equation}\label{AntwoE}
\begin{array}{l}
u-v=0, \quad x \in \partial \mathcal O, \\
\frac{\partial u}{\partial \nu} - \frac{\partial v}{\partial \nu_A}=0, \quad x \in \partial \mathcal O,
\end{array}
\end{equation}
where $\lambda=k^2,~~\psi_{\text{sc}}$ satisfies the radiation conditions: $$
\psi_{\text{sc}}=f(k,\theta,\omega)\frac{e^{ikr}}{r^{(d-1)/2}}+O \left (\frac{1}{r^{(d+1)/2}} \right ),\quad \theta=\frac{x}{r},~~r=|x|\to\infty.
$$
\begin{proposition}
If $z=1$ is an eigenvalue of the scattering matrix $S_{\text{tr}}(k),~k>0,$ for the transmission scattering problem (\ref{AnoneE}), (\ref{AntwoE}), then $k^2$ is one of the ITE.
\end{proposition}
The proof of this statement is the same as in the case of the Dirichlet or Neumann boundary conditions. The inverse statement is also valid for the transmission problem when $\mathcal O$ is a ball, $A=a~\!I\!d, a$ and $n$ are constant. We believe that the result on the weak implication from B to A also holds for the transmission problem (for arbitrary domains $\mathcal O$), but it has not been proved yet.

Let us recall conditions on $A,n$ which guarantee the parameter-ellipticity of interior transmission problem (\ref{Anone}), (\ref{Antwo}). In this paper we assume that $A$ and $n$ are real-valued. Let us fix an arbitrary point $x^0\in \partial \mathcal O $ and choose a new orthonormal basis $\{e_j\},~1\leq j\leq d,$ centered at the point $x^0$ with $e_d=\nu$, where $\nu$ is the normal to the boundary at the point $x^0$.  The vectors $e_1,...,e_{d-1}$ belong to the tangent plane to $\partial \mathcal O $ at the point $x_0$. Let $y$ be the local coordinates defined by the basis $\{e_j\}$, and let $C=C(x^0)$ be the transfer matrix, i.e., $y=C(x-x^0)$.

We fix the point $x=x^0$ in  equations (\ref{Anone}), (\ref{Antwo}) and rewrite the problem in the local coordinates $y$. Then we get the following problem with constant coefficients in the half space $y_d>0:$
\begin{equation}\label{Anone1}
\begin{array}{l}
-\Delta_y u - k^2 u =0, \quad y_d>0 ,\\
 -\nabla_y \widetilde{A} \nabla_y v - k^2 n(x^0)v =0,  \quad y_d>0,
\end{array}
\end{equation}
\begin{equation}\label{Antwo1}
\begin{array}{l}
u-v=0, \quad y_d=0, \\
\frac{\partial u}{\partial {y_d}} - \frac{\partial v}{\partial \nu_{\widetilde{A}}}=0, \quad y_d=0.
\end{array}
\end{equation}
Here
$$
\widetilde{A}=\widetilde{A}(x^0)=CA(x^0)C^*.
$$
The entries of the matrix $\widetilde{A}=(a_{i,j})$ are equal to $a_{i,j}=e_j\cdot A(x^0)e_i$. The co-normal derivative in the boundary condition equals $e_d\cdot \widetilde{A}\nabla_y.$

The following result can be extracted from \cite{LV4}.
\begin{theorem}\label{a3} Let $A(x)>0, n(x)>0$ for $  x \in \mathcal O$ and the following conditions hold for all $  x^0 \in \partial\mathcal O$: if $d=2$, then
\[
a_{2,2}n(x^0)-1 \neq 0,~~{\rm {and}}~~ {\rm{det}} A(x^0)\neq 1,\quad  x^0  \in \partial \mathcal O;
\]
if $d=3$, then $a_{3,3}n(x^0)-1 \neq 0$ and
$$
det \left ( \begin{array}{ll} a_{3,3}a_{1,1}-(a_{1,3})^2 -1 & a_{3,3}a_{1,2}-a_{1,3}a_{2,3} \\
a_{3,3}a_{2,1}-a_{1,3}a_{2,3} & a_{3,3}a_{2,2}-(a_{2,3})^2 -1 \end{array} \right ) >0, \quad x^0\in \partial \mathcal O.
$$

Then the interior transmission problem (\ref{Anone}), (\ref{Antwo}) is parameter-elliptic for $\lambda$ in any sector of the complex $\lambda$-plane which does not contain either of the rays $R_+$ and $R_-$. Its eigenvalues form a discrete set.

Moreover, if $\sigma=1$, then the problem is parameter-elliptic also on $R_-$. Here
$$\sigma=\rm{sgn}[(a_{2,2}n(x^0)-1 )({\rm{det}} A(x^0)-1)], \quad  x^0  \in \partial \mathcal O,
$$
if $d=2$. If $d=3$, then $\sigma=1$ when the matrix
$$
\left ( \begin{array}{lll} a_{3,3}n(x^0)-1 & 0& 0\\0&  a_{3,3}a_{1,1}-(a_{1,3})^2 -1 & a_{3,3}a_{1,2}-a_{1,3}a_{2,3} \\
0& a_{3,3}a_{2,1}-a_{1,3}a_{2,3} & a_{3,3}a_{2,2}-(a_{2,3})^2 -1 \end{array} \right ), \quad x^0\in \partial \mathcal O,
$$
is sign-definite and $\sigma=-1$ otherwise.
\end{theorem}

\section{Weyl asymptotics and branching billiards.}

Let us split the set $\{\lambda\},~\lambda \in \mathbb C,$ of ITE into two subsets  $\{\lambda^+\}$ and $\{\lambda^-\}$ where Re$\lambda^+\geq 0$, Re$\lambda^-< 0$. We enumerate the ITE $\lambda^+_n,\lambda^-_n$ in increasing order of $|\lambda_n|$ and denote by $N^+(t),N^-(t)$ the counting functions for $\{\lambda^+_n\}$,$\{\lambda^-_n\}$. Similarly, we enumerate the whole set of the ITE in increasing order of $|\lambda_n|$ and denote by $N(t)$ the counting functions for $\{\lambda_n\}$.
\begin{theorem}\label{teorema1}Let assumptions of Theorem \ref{a3} hold (with arbitrary $\sigma=\pm 1$). Then

1)
There is at most a finite number of ITE inside any sector of complex $\lambda$-plane which does not contain either of the rays $R_+$ and $R_-$.

2) The Weyl formula holds for $\{\lambda^+_n\}$:
\begin{equation}\label{est1}
N^+(t) \sim \alpha t^{d/2}, \quad t \rightarrow \infty,
\end{equation}
where
\begin{equation}\label{est2}
\alpha=\frac{\omega_d}{(2\pi)^d} \int_{\mathcal O} \left (1+\frac{n^{d/2}(x)}{(det A(x))^{1/2}} \right ) dx.
\end{equation}
In particular, if  $A=a I\!d$, where $a>0$, and $n$ is also constant, then
$$
\alpha=\frac{\omega_d}{(2\pi)^d} V(\mathcal O) \left (1+ \left (\frac{n}{a} \right)^{d/2} \right ).
$$

3) If $\sigma=1$ then the set $\{\lambda^-_n\}$ is finite. If $\sigma=-1$ then there exists $M>0$ such that
$$
N^-(t) <M t, \quad t \rightarrow \infty, \quad d=2; \quad N^-(t) <M t\ln t, \quad t \rightarrow \infty, \quad d=3.
$$
\end{theorem}

\textbf{Remark} If
 $n(x)=a, A(x) = a I\!d $, where $a>0, a \neq 1$, then the substitution $u-v=u_1,~ au-v=v_1$ reduces the problem  (\ref{Anone}),(\ref{Antwo}) to the Dirichlet problem for $u_1$ and the Neumann problem for $v_1$. Hence in this case, the set $\{\lambda_n\}$ coincides with the union of the Dirichlet and Neumann Laplacians.

\textbf{Proof.} The first statement of the theorem and the first part of the last statement (concerning $\sigma=1$) are immediate consequences of the parameter-ellipticity of the problem established in Theorem \ref{a3}. The second statement follows from Theorem \ref{a3} and results obtained in \cite{BK},\cite{BK2} where the main term of the Weyl asymptotics is justified for non-symmetric parameter-elliptic system if the eigenvalues of the main symbol $\mathcal A$ of the system belong to $R_+$. Formula (\ref{est1}) holds in this case with
$$
\alpha=\frac{1}{(2\pi)^{d}}\int_{\mathcal O} \int_{R^d} N(1,x,\xi) dx d\xi,
$$
where $N(1,x,\xi)$ is the number of eigenvalues of $\mathcal A$ whose absolute values do not exceed one. In our case,
$$
\mathcal A=\left (
\begin{array}{cc}
\xi^2 & 0 \\
0 & \frac{1}{n(x)}\xi^t A(x) \xi  \\
\end{array} \right ),
$$
and
$$
\alpha=\frac{1}{(2\pi)^{d}}\int_{\mathcal O} dx \left ( \int_{|\xi|^2<1} +\int_{\xi^t A \xi <n(x)} \right ) d\xi,
$$
which implies (\ref{est2}).

The part of the third statement concerning the case $\sigma=-1$ is a consequence of Theorem 2 from \cite{BK},\cite{BK2}. The latter theorem estimates the number of the eigenvalues of the problem in a sector, which does not contain the eigenvalues of the principle symbol of the operator.
\qed

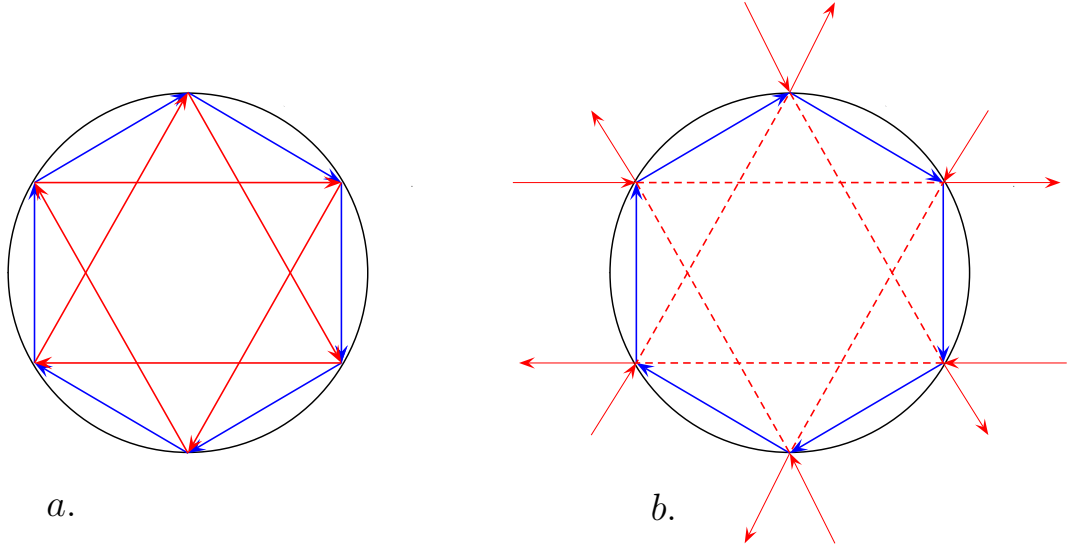
\begin{figure}[h]
\begin{picture}(0,210)

\rput(3.0,3.8){
\scalebox{2.4}{
\psdots[dotsize=0.15pt](-1,0)(0.534,0.925)(1.871,1.732)(1.241,0.479)
\rput(-0.7,-1.3){\scalebox{0.5}{$a.$}}
\pspolygon[linewidth=0.03pt,linecolor=white]
(0.534,0.925)(1.871,1.732)(1.64,1.483)(1.48,1.304)(1.34,1.14)(1.223,1)(1.18,0.947)
(1.136,0.89)(1.093,0.8367)(1.0456,0.7746)(1,0.709)(0.9,0.773)(0.8,0.826)(0.7,0.87)(0.6,0.906)
\pspolygon[linewidth=0.03pt,linecolor=white]
(0.534,-0.925)(1.871,-1.732)(1.64,-1.483)(1.48,-1.304)(1.34,-1.14)(1.223,-1)(1.18,-0.947)
(1.136,-0.89)(1.093,-0.8367)(1.0456,-0.7746)(1,-0.709)(0.9,-0.773)(0.8,-0.826)(0.7,-0.87)(0.6,-0.906)
\psellipse[linewidth=0.233pt](0,0)(1,1)

  \psline[linewidth=0.25pt,linecolor=blue, arrows=<-,arrowscale=1]
  (-0.85,0.5)(-0.85,-0.5)

  \psline[linewidth=0.25pt,linecolor=blue, arrows=<-,arrowscale=1]
  (0.85,-0.5)(0.85,0.5)

  \psline[linewidth=0.25pt,linecolor=blue, arrows=<-,arrowscale=1]
  (0.85,0.5)(0,1)

  \psline[linewidth=0.25pt,linecolor=blue, arrows=<-,arrowscale=1]
  (0,1)(-0.85,0.5)

  \psline[linewidth=0.25pt,linecolor=blue, arrows=<-,arrowscale=1]
(-0.85,-0.5)(0,-1)

  \psline[linewidth=0.25pt,linecolor=blue, arrows=<-,arrowscale=1]
  (0,-1)(0.85,-0.5)

  \psline[linewidth=0.25pt,linecolor=blue, arrows=<-,arrowscale=1]
  (0.85,-0.5)(0.85,-0.5)

  \psline[linewidth=0.25pt,linecolor=red, arrows=<-,arrowscale=1]
  (0.85,-0.5)(0,1)

  \psline[linewidth=0.25pt,linecolor=red, arrows=<-,arrowscale=1]
(0,1)(-0.85,-0.5)

  \psline[linewidth=0.25pt,linecolor=red, arrows=<-,arrowscale=1]
  (-0.85,-0.5)(0.85,-0.5)

  \psline[linewidth=0.25pt,linecolor=red, arrows=<-,arrowscale=1]
  (-0.85,0.5)(0,-1)

  \psline[linewidth=0.25pt,linecolor=red, arrows=<-,arrowscale=1]
  (0,-1)(0.85,0.5)

  \psline[linewidth=0.25pt,linecolor=red, arrows=<-,arrowscale=1]
  (0.85,0.5)(-0.85,0.5)

}}

\rput(11.0,3.8){
\scalebox{2.4}{
\psdots[dotsize=0.15pt](-1,0)(0.534,0.925)(1.871,1.732)(1.241,0.479)
\rput(-0.7,-1.3){\scalebox{0.5}{$b.$}}
\pspolygon[linewidth=0.03pt,linecolor=white]
(0.534,0.925)(1.871,1.732)(1.64,1.483)(1.48,1.304)(1.34,1.14)(1.223,1)(1.18,0.947)
(1.136,0.89)(1.093,0.8367)(1.0456,0.7746)(1,0.709)(0.9,0.773)(0.8,0.826)(0.7,0.87)(0.6,0.906)
\pspolygon[linewidth=0.03pt,linecolor=white]
(0.534,-0.925)(1.871,-1.732)(1.64,-1.483)(1.48,-1.304)(1.34,-1.14)(1.223,-1)(1.18,-0.947)
(1.136,-0.89)(1.093,-0.8367)(1.0456,-0.7746)(1,-0.709)(0.9,-0.773)(0.8,-0.826)(0.7,-0.87)(0.6,-0.906)
\psellipse[linewidth=0.233pt](0,0)(1,1)

  \psline[linewidth=0.25pt,linecolor=blue,arrows=->,arrowscale=1]
  (0.85,0.5)(0.85,-0.5)

  \psline[linewidth=0.25pt,linecolor=blue,arrows=->,arrowscale=1]
(0,1)(0.85,0.5)

  \psline[linewidth=0.25pt,linecolor=blue,arrows=->,arrowscale=1]
(-0.85,0.5)(0,1)

  \psline[linewidth=0.25pt,linecolor=blue,arrows=->,arrowscale=1]
(-0.85,-0.5)(-0.85,0.5)

  \psline[linewidth=0.25pt,linecolor=blue,arrows=->,arrowscale=1]
(0,-1)(-0.85,-0.5)

  \psline[linewidth=0.25pt,linecolor=blue,arrows=->,arrowscale=1]
(0.85,-0.5)(0,-1)

  \psline[linewidth=0.25pt,linestyle=dashed,dash=1.2pt 0.8pt,linecolor=red]
  (0.85,-0.5)(0,1)(-0.85,-0.5)(0.85,-0.5)

  \psline[linewidth=0.15pt,linecolor=red,arrows=->,arrowscale=1]
 (-0.85,0.5)(-1.1,0.9)

  \psline[linewidth=0.15pt,arrows=->,linecolor=red,arrowscale=1]
 (0.85,-0.5)(1.1,-0.9)

  \psline[linewidth=0.15pt,linecolor=red,arrows=->,arrowscale=1]
 (1.1,0.9)(0.85,0.5)

\psline[linewidth=0.15pt,linecolor=red,arrows=->,arrowscale=1]
 (-1.1,-0.9)(-0.85,-0.5)

  \psline[linewidth=0.15pt,linecolor=red,arrows=->,arrowscale=1]
(-1.534,0.5)(-0.85,0.5)

 \psline[linewidth=0.15pt,linecolor=red,arrows=->,arrowscale=1]
(1.534,-0.5)(0.85,-0.5)

  \psline[linewidth=0.15pt,linecolor=red,arrows=->,arrowscale=1]
(0.85,0.5)(1.5,0.5)

  \psline[linewidth=0.15pt,linecolor=red,arrows=->,arrowscale=1]
(-0.85,-0.5)(-1.5,-0.5)

  \psline[linewidth=0.15pt,linecolor=red,arrows=->,arrowscale=1]
(-0.25,1.5)(0,1)

  \psline[linewidth=0.15pt,linecolor=red,arrows=->,arrowscale=1]
(0,1)(0.25,1.5)

  \psline[linewidth=0.15pt,linecolor=red,arrows=->,arrowscale=1]
(0.25,-1.5)(0,-1)

  \psline[linewidth=0.15pt,linecolor=red,arrows=->,arrowscale=1]
(0,-1)(-0.25,-1.5)

  \psline[linewidth=0.25pt,linestyle=dashed,dash=1.2pt 0.8pt,linecolor=red]
  (-0.85,0.5)(0,-1)(0.85,0.5)(-0.85,0.5)

}}

\end{picture}
\caption{ Strongly periodic branching trajectory of the interior problem and the corresponding trajectory of the exterior problem. }\label{fig3}
\end{figure}

It is well known that the remainder term in the classical Weyl asymptotics depends on the set of the closed billiard trajectories (periodic trajectories of the corresponding Hamiltonian system). The second term of the asymptotics was justified by V. Ivrii \cite{Ivr} for boundary value problems for the Laplacian under the condition that the set of periodic trajectories has zero measure.
D. Vasiliev \cite{vas} and Yu. Safarov \cite{saf} extended this result to the case of self-adjoint elliptic system where the billiard trajectories are not defined uniquely after a ray hits the boundary (see figure \ref{fig3}a). We can not use the latter results directly since the problem (\ref{Anone}),(\ref{Antwo}) is not symmetric, but we will combine these results with the theory of $s$-numbers to obtain an estimate for the second term of the asymptotics from above. Let us describe the branching trajectories for our problem.

Consider two Hamiltonian flows $G_i^t,~i=1,2,$ with Hamiltonians
\begin{equation}\label{defham}
h_1(x,\xi)=|\xi|, \quad h_2(x,\xi)=\sqrt{\frac{1}{n(x)} \xi^t A(x) \xi}, \quad x \in \mathcal O,~ \xi \in R^d ~~(= T'(\mathcal O)).
\end{equation}

We assume that $h_1(x,\xi) \neq h_2(x,\xi), \forall x \in \mathcal O, \xi \in R^d\backslash \{0\}$, which is equivalent to the following:
\begin{assumption}\label{ass3}
Matrix
$\frac{1}{n(x)} A-I\!d$ is positive-definite or negative-definite when $x \in \overline{\mathcal O}$.
\end{assumption}

When a ray which corresponds to one of these Hamiltonians comes to the boundary $\partial\mathcal O$ it creates two reflected rays. One of the reflection angles corresponds to the "billiard law": reflection angle is equal to the incident angle. This ray corresponds to the same Hamiltonian as the Hamiltonian of the incident ray. The second reflection ray is defined buy Snell's law. One can determine both reflected rays as follows. If the Hamiltonian of the incident ray is equal to one (it is constant along  the trajectory), then the directions of the reflected rays are defined by the relations $h_1=h_2=1$ at the point of reflection (and it may happen that one of these equations does not have a solution in which case there is only one reflected ray).

One needs to work with branching trajectories in order to construct quasi-modes (approximate solutions of the problem). A branching trajectory splits every time when it has a chance to split after the reflection from the boundary. After the initial Hamiltonian $h_{i_0}$ (to start the trajectory) is chosen, the branching trajectory is defined uniquely by its initial data $(y,\eta),~y\in\mathcal O, ~ \eta \in R^d,~h_{i_0}(y,\eta)=1$. The initial data form a $(2d-1)$-dimensional manifold. For our purpose (an estimate on the second term in the Weyl asymptotics), we need another object, billiard trajectories. These trajectories $(x^t,\xi^t),~t\geq 0,$ do not split. The point moves according to one of the Hamiltonian flows and the Hamiltonian can be changed after each reflection from the boundary. Thus each initial data usually defines infinitely many billiard trajectories.

A billiard trajectory is called a dead-end trajectory if the ray touches the boundary or there are infinitely many reflections on a finite time interval.

\begin{assumption}\label{ass1}
The measure of the dead-end trajectories (i.e., the $(2d-1)$-Lebesgue measure of the set of initial data for the dead-end trajectories) is zero.
\end{assumption}

Consider a periodic billiard trajectory with a period $T$ and initial data $(y_0,\eta_0)$. This trajectory is called absolutely periodic if for each $(y,\eta)$ in a $\rho$-neighborhood of $(y_0,\eta_0)$ a billiard trajectory  $(x^t,\xi^t),~t\geq 0,$ with the initial data $(y,\eta)$ can be chosen in such a way (by repeating the same pattern of reflections) that the trajectory at time $t=T$ is located in $O(\rho^\infty),~\rho\to 0,$ neighborhood of the starting point $(y,\eta)$. The latter means that the function $(x^t,\xi^t)-(y,\eta)$ has zero of infinite order at the point $(y_0,\eta_0)$. The initial data $(y_0,\eta_0)$ is called absolutely periodic if at least one absolutely periodic trajectory has this data.
\begin{assumption}\label{ass2a}
 $(2d-1)$-Lebesgue measure of absolutely periodic data is zero.
\end{assumption}
\begin{theorem}\label{teor2}
Let the assumptions of Theorem \ref{a3} and Assumptions \ref{ass3}-\ref{ass2a} hold. Then there is a constant $C$ such that
\begin{equation}\label{kaknado}
N(t) \leq \alpha t^{d/2} + C t^{(d-1)/2} + o(t^{(d-1)/2}), \quad t \rightarrow \infty,
\end{equation}
where $\alpha$ is defined in (\ref{est2}).
\end{theorem}
\textbf{Remark.} Assumption \ref{ass2a} can be weakened, see \cite{saf}.

\textbf{Proof.} Denote by $L^2_n$ the weighted space $L^2(\mathcal O)\times L^2(\mathcal O)$ with the weight which corresponds to the following scalar product:
\begin{equation}\label{dotpr}
(u_1,v_1)\cdot(u_2,v_2)=\int_{\mathcal O} u_1 \overline{u}_2 dx + \int_{\mathcal O} n(x) v_1 \overline{v}_2 dx.
\end{equation}
Consider the following operator $L$ in $L^2_n$ which corresponds to problem  (\ref{Anone}), (\ref{Antwo}): the mapping  $L$ is defined by
\begin{equation}\label{1234}
L(u,v)=(-\Delta u,\frac{-1}{n(x)}\nabla A(x)\nabla v),
\end{equation}
and the domain $D_L$ consists of vectors $(u,v),~u,v\in H^2(\mathcal O),$ such that $u,v$ satisfy (\ref{AntwoE}).
Operator $L^*$, adjoint to $L$ with respect to the scalar product (\ref{dotpr}), can be obtained from $L$ if we replace the minus signs in both relations (\ref{AntwoE}) by the plus signs.

Let us fix a real $\gamma$ which is not an ITE, i.e., the operators $(L-\gamma I)$ and $(L^*-\gamma I)$ are invertible.
Consider the operator $M=(L-\gamma I)^*(L-\gamma I)$ defined on $(u,v), u,v \in H^4(\mathcal O),$ such that $(u,v)\in D_L,~(L-\gamma)(u,v)\in D_{L^*}$. Operator $M$ in $L_n^2$ with the domain $D_M$ described above is self-adjoint. One can rewrite the equation $Mz=\lambda z, ~z=(u,v)$, in the form of the boundary value problem:
\begin{equation}\label{1AnoneE}
\begin{array}{l}
(-\Delta-\gamma)^2 u = \lambda u, \quad x \in \mathcal O, \\
(\frac{-1}{n(x)}\nabla A \nabla -\gamma)^2  v = \lambda v , \quad x \in \mathcal O,
\end{array}
\end{equation}
\begin{equation}\label{1AntwoE}
\begin{array}{l}
u-v=0, \quad x \in \partial \mathcal O, \\
\frac{\partial u}{\partial \nu} - \frac{\partial v}{\partial \nu_A}=0, \quad x \in \partial \mathcal O,\\
(\Delta+\gamma)u+(\frac{1}{n(x)}\nabla A \nabla+\gamma)v=0, \quad x \in \partial \mathcal O, \\
\frac{\partial (\Delta+\gamma) u}{\partial \nu} +\frac{\partial  (\frac{1}{n(x)}\nabla A \nabla+\gamma) v}{\partial \nu_{A}}=0, \quad x \in \partial \mathcal O.
\end{array}
\end{equation}
If the assumptions of Theorem \ref{a3} hold, then problem (\ref{1AnoneE}), (\ref{1AntwoE}) is elliptic. Indeed, one can check that the Shapiro-Lopatinskii condition for problem (\ref{1AnoneE}), (\ref{1AntwoE}) coincides with the ones for the problem  (\ref{Anone}), (\ref{Antwo}). Another option is to note that the standard elliptic a priory estimates are valid for the solutions of inhomogeneous problem (\ref{1AnoneE}), (\ref{1AntwoE}) with inhomogeneities in  (\ref{1AnoneE}) (since they are valid for operators $(L-\gamma I)$ and $(L^*-\gamma I)$) and the latter implies the ellipticity of the boundary value problem. Since problem (\ref{1AnoneE}), (\ref{1AntwoE})
is symmetric, we can apply the results of  D. Vassiliev \cite[Th.1.2]{vas} and Yu. Safarov \cite[Th.1.1]{saf} which provide an estimate on the second term of the Weyl asymptotics for symmetric elliptic systems. Namely, it follows that if the assumptions of Theorem \ref{a3} and Assumptions \ref{ass3}-\ref{ass2a} hold, then
\begin{equation}\label{noft}
\widehat{N}(t)=\alpha t^{\frac{d}{4}}+C t^{\frac{d-1}{4}}+o(t^{\frac{d-1}{4}}), \quad t \rightarrow \infty,
\end{equation}
where $\widehat{N}(t)$ is the counting function for operator $(L-\gamma I)^*(L-\gamma I)$ and $\alpha$ is defined in (\ref{est2}).
 Operators
$[(L-\gamma I)^*(L-\gamma I)]^{-1}$ and $(L-\gamma I)^{-1}$ are compact, and therefore (e.g., see \cite[Lemma 3.3]{gk})
\begin{equation}\label{lam1}
|\lambda_n((L-\gamma I)^{-1})|^2 \leq \lambda_n([(L-\gamma I)^*(L-\gamma I)]^{-1}),
\end{equation}
where the eigenvalues $\lambda_n$ are enumerated in the order of the decay of $|\lambda_n|$ (contrary to the eigenvalues of operators $M$ or $L$ for which $|\lambda_n|\to\infty$ as $n\to\infty$). Inequality (\ref{lam1}) implies that
$$
|\lambda_n((L-\gamma I))|^2 \geq \lambda_n([(L-\gamma I)^*(L-\gamma I)]),
$$
and this together with (\ref{noft}) justify the statement of the theorem.
\qed

\section{Strongly periodic branching trajectories.}\label{bsv}

\begin{figure}[h]
\begin{picture}(0,210)

\rput(12.0,3.8){
\scalebox{2.2}{
\psdots[dotsize=0.15pt](-1,0)(0.534,0.925)(1.871,1.732)(1.241,0.479)
\rput(-0.7,-1.3){\scalebox{0.5}{$b.$}}
\pspolygon[linewidth=0.03pt,linecolor=white]
(0.534,0.925)(1.871,1.732)(1.64,1.483)(1.48,1.304)(1.34,1.14)(1.223,1)(1.18,0.947)
(1.136,0.89)(1.093,0.8367)(1.0456,0.7746)(1,0.709)(0.9,0.773)(0.8,0.826)(0.7,0.87)(0.6,0.906)
\pspolygon[linewidth=0.03pt,linecolor=white]
(0.534,-0.925)(1.871,-1.732)(1.64,-1.483)(1.48,-1.304)(1.34,-1.14)(1.223,-1)(1.18,-0.947)
(1.136,-0.89)(1.093,-0.8367)(1.0456,-0.7746)(1,-0.709)(0.9,-0.773)(0.8,-0.826)(0.7,-0.87)(0.6,-0.906)
\psellipse[linewidth=0.233pt](0,0)(1,1)


  \psline[linewidth=0.35pt,linecolor=blue,arrows=->,arrowscale=1]
  (0,-1)(-0.85,0.5)

  \psline[linewidth=0.35pt,linecolor=blue,arrows=->,arrowscale=1]
  (0.85,0.5)(0,-1)

  \psline[linewidth=0.35pt,linecolor=blue,arrows=->,arrowscale=1]
  (-0.85,0.5)(0.85,0.5)

  \psline[linewidth=0.25pt,linecolor=red,arrows=->,arrowscale=1]
 (1.0,1.4)(0.85,0.5)
  \psline[linewidth=0.25pt,linestyle=dashed,dash=1.2pt 0.8pt,linecolor=red]
(0.85,0.5)(0.6, -1)

  \psline[linewidth=0.25pt,linestyle=dashed,dash=1.2pt 0.8pt,linecolor=red]
(0.85,0.5)(-1, 1.2)

  \psline[linewidth=0.25pt,linecolor=red,arrows=->,arrowscale=1]
(0.85,0.5) (1.6,0.1)


 \psline[linewidth=0.25pt,linecolor=red,arrows=->,arrowscale=1]
 (0.5,-1.5)(0,-1)

 \psline[linewidth=0.25pt,linestyle=dashed,dash=1.2pt 0.8pt,linecolor=red]
(0, -1) (1,-0.1)

 \psline[linewidth=0.25pt,linestyle=dashed,dash=1.2pt 0.8pt,linecolor=red]
(0, -1) (-1,-0.1)

  \psline[linewidth=0.25pt,linecolor=red,arrows=->,arrowscale=1]
  (0, -1)(-0.65,-1.5)


  \psline[linewidth=0.25pt,linecolor=red,arrows=->,arrowscale=1]
 (-0.84,0.5)(-1.0,1.5)

  \psline[linewidth=0.25pt,linestyle=dashed,dash=1.2pt 0.8pt,linecolor=red]
(-0.85,0.5)(-0.6, -1)

  \psline[linewidth=0.25pt,linestyle=dashed,dash=1.2pt 0.8pt,linecolor=red]
(-0.85,0.5)(1, 1.2)

  \psline[linewidth=0.25pt,linecolor=red,arrows=->,arrowscale=1]
  (-1.6,0.1)(-0.85,0.5)

}}

\rput(3.5,3.8){
\scalebox{2.2}{
\psdots[dotsize=0.15pt](-1,0)(0.534,0.925)(1.871,1.732)(1.241,0.479)
\rput(-0.7,-1.3){\scalebox{0.5}{$a.$}}
\pspolygon[linewidth=0.03pt,linecolor=white]
(0.534,0.925)(1.871,1.732)(1.64,1.483)(1.48,1.304)(1.34,1.14)(1.223,1)(1.18,0.947)
(1.136,0.89)(1.093,0.8367)(1.0456,0.7746)(1,0.709)(0.9,0.773)(0.8,0.826)(0.7,0.87)(0.6,0.906)
\pspolygon[linewidth=0.03pt,linecolor=white]
(0.534,-0.925)(1.871,-1.732)(1.64,-1.483)(1.48,-1.304)(1.34,-1.14)(1.223,-1)(1.18,-0.947)
(1.136,-0.89)(1.093,-0.8367)(1.0456,-0.7746)(1,-0.709)(0.9,-0.773)(0.8,-0.826)(0.7,-0.87)(0.6,-0.906)
\psellipse[linewidth=0.233pt](0,0)(1,1)


  \psline[linewidth=0.35pt,linecolor=blue,arrows=->,arrowscale=1]
  (0,-1)(-0.85,0.5)

  \psline[linewidth=0.35pt,linecolor=blue,arrows=->,arrowscale=1]
  (0.85,0.5)(0,-1)

  \psline[linewidth=0.35pt,linecolor=blue,arrows=->,arrowscale=1]
  (-0.85,0.5)(0.85,0.5)

  \psline[linewidth=0.25pt,linestyle=dashed,dash=1.2pt 0.8pt,linecolor=red]
(0.85,0.5)(0.6, -1)

  \psline[linewidth=0.25pt,linestyle=dashed,dash=1.2pt 0.8pt,linecolor=red]
(0.85,0.5)(-1, 1.2)

  \psline[linewidth=0.25pt,linecolor=red,arrows=->,arrowscale=1]
(0.85,0.5) (1.6,0.1)



 \psline[linewidth=0.25pt,linestyle=dashed,dash=1.2pt 0.8pt,linecolor=red]
(0, -1) (1,-0.1)

 \psline[linewidth=0.25pt,linestyle=dashed,dash=1.2pt 0.8pt,linecolor=red]
(0, -1) (-1,-0.1)

  \psline[linewidth=0.25pt,linecolor=red,arrows=->,arrowscale=1]
  (0, -1)(-0.65,-1.5)


  \psline[linewidth=0.25pt,linecolor=red,arrows=->,arrowscale=1]
 (-0.84,0.5)(-1.0,1.5)

  \psline[linewidth=0.25pt,linestyle=dashed,dash=1.2pt 0.8pt,linecolor=red]
(-0.85,0.5)(-0.6, -1)

  \psline[linewidth=0.25pt,linestyle=dashed,dash=1.2pt 0.8pt,linecolor=red]
(-0.85,0.5)(1, 1.2)

  \psline[linewidth=0.25pt,linecolor=red,arrows=->,arrowscale=1]
  (-1.6,0.1)(-0.85,0.5)

}}

\end{picture}
\caption{ Periodic branching trajectories of the exterior transmission problem which are not related to the periodic branching trajectories of the interior problem. There is only one incident ray in a), and three incident rays in b). }\label{fig2}
\end{figure}
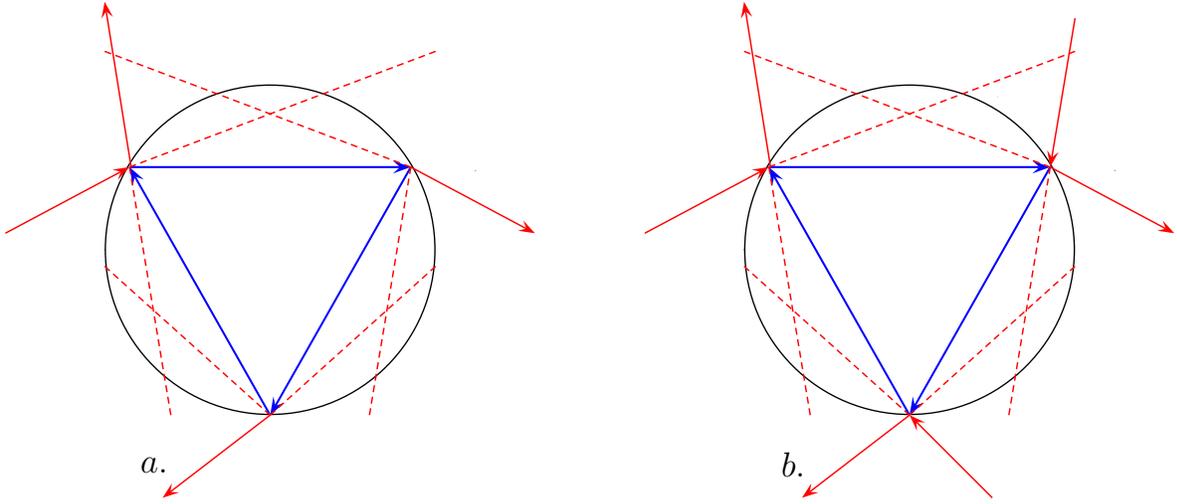

Consider again the branching billiard trajectories of the ITE problem. In the previous section, we considered periodic billiard trajectories where the point moves (after hitting the boundary) along one of the possible reflected rays. Now we assume that the trajectory splits every time when it has a chance to split. We will call such a trajectory strongly periodic branching billiard trajectory if it consists of a finite number of segments (see Figure \ref{fig3}a). Under some additional condition on the optical lengths of some parts of the trajectory, one can use these trajectories to construct quasi-modes for the ITE problem, i.e., to construct functions which almost (as $k=k_n\to\infty$) satisfy the equations and boundary conditions of the ITE problem.

One also can construct quasi-modes for the exterior transmission problem. The corresponding billiard trajectories are defined by the Hamiltonian flow outside of the obstacle with the Hamiltonian $h_1(x,\xi)=|\xi|$ and the Hamiltonian flow inside of the obstacle with the Hamiltonian $h_2(x,\xi)=\sqrt{\frac{1}{n(x)} \xi^t A(x) \xi}.$ After hitting the boundary, each ray splits into a reflected and refracted parts. Such a trajectory is called periodic if it consists of a finite number of segments (exterior rays can be semi-infinite, see Figures \ref{fig3}b and \ref{fig2}).

While the periodic trajectories of the exterior problem may not be related to strongly periodic trajectories of the interior problem (Figure \ref{fig2}), each strongly periodic branching trajectory of the interior problem always define a corresponding trajectory of the exterior problem. In order to define such exterior trajectory, one needs to replace each segment of the interior trajectory which corresponds to the Hamiltonian  $h_1(x,\xi)=|\xi|$ by its complement on the line containing this segment (see Figure \ref{fig3}). One also can extend to $R^d\backslash\mathcal O$ the parts of the interior quasi-mode which correspond to $h_1$, and then we omit the interior parts of the quasi-mode related to $h_1$. This will provide a quasi-mode of the exterior problem which has the following property. Its exterior part can be smoothly extended inside of the obstacle as a solution of the Helmholtz equation in the whole space (where the extension is the omitted part of the interior mode). The latter is an important feature in the relation between the ITE and the eigenvalues of the $S$-matrix. In particular, this feature provides an alternative proof that property B implies property A which is discussed in the first section.

\textbf{Acknowledgment.} The authors are very grateful to F. Cakoni, D. Colton, B. Gutkin, H. Haddar, V. Ivrii, A. Kirsch,  Yu. Safarov, D. Vassiliev  for useful discussions.

\end{document}